\documentclass[prb,twocolumn,superscriptaddress,floatfix]{revtex4-2}
\usepackage{amsfonts}
\usepackage{graphicx}
\usepackage{float}
\usepackage{amsmath}
\usepackage{amssymb}
\usepackage{hyperref}
\usepackage{pgf}
\usepackage{physics}
\usepackage[left=0.5in,right=0.5in,top=0.7in,bottom=0.7in
]{geometry}
\begin{document}

\title{Efficient optimization of variational tensor-network approach to
three-dimensional statistical systems}
\author{Xia-Ze Xu}
\thanks{These authors contributed equally.}
\affiliation{State Key Laboratory of Low Dimensional Quantum Physics and Department of
Physics, Tsinghua University, Beijing 100084, China}

\author{Tong-Yu Lin}
\thanks{These authors contributed equally.}
\affiliation{State Key Laboratory of Low Dimensional Quantum Physics and Department of
Physics, Tsinghua University, Beijing 100084, China}

\author{Guang-Ming Zhang}
\email{zhanggm@shanghaitech.edu.cn}
\affiliation{School of Physical Science and Technology, ShanghaiTech University, Shanghai
201210, China}
\affiliation{Department of Physics, Tsinghua University,
Beijing 100084, China}

\date{\today }

\begin{abstract}
Variational tensor network optimization has become a powerful tool for
studying classical statistical models in two dimensions. However, its
application to three-dimensional systems remains limited, primarily due to
the high computational cost associated with evaluating the free energy
density and its gradient. This process requires contracting a triple-layer
tensor network composed of a projected entangled pair operator and projected
entangled pair states. In this paper, we employ a split corner-transfer
renormalization group scheme tailored for the contraction of such a
triple-layer network, which reduces the computational complexity while
keeping high accuracy. Through numerical benchmarks on the three-dimensional
classical Ising model, we demonstrate that the proposed scheme achieves
numerical results comparable to the most recent Monte Carlo simulations,
providing a substantial speedup over previous variational tensor network
approaches. This makes this method well-suited for efficient gradient-based
optimization in three-dimensional tensor network simulations.
\end{abstract}

\maketitle

\section{Introduction}

The study of statistical models significantly deepens our understanding of
collective phenomena and phase transitions. However, exactly solvable models
remain rare, leading contemporary research to rely increasingly on numerical
methods such as Monte Carlo simulations \cite%
{talapov1996,feng_2010,ferrenberg_2018,Hasenbusch2021} and the conformal
bootstrap \cite{showk_2012,showk_2014,kos_2016}. In last decades, tensor
network have emerged as a powerful numerical framework for studying
statistical models. In this approach, the partition function is represented
as a tensor network and then approximately contracted. For two-dimensional
(2D) statistical models, tensor network methods are well-established, with
techniques like the imaginary time evolution \cite{vidal_2007,orus_2008},
the tensor renormalization group (TRG) \cite%
{levin_2007,zhao_2010,xie_2009,xie_2012}, the corner transfer matrix
renormalization group (CTMRG) \cite{nishino_1996}, and the gradient-based
variational method demonstrating both high efficacy and accuracy \cite%
{haegeman_2017,zauner_2018,vanderstraeten_2019} These methods have proven
efficient in addressing a broad range of problems, including spin models
with both discrete and continuous degrees of freedom \cite%
{vanderstraeten_2019_2,Li_2020,schmoll_2021}, as well as both non-frustrated
and frustrated spin systems \cite{vanhecke_2021,colbois_2022a,song2023a}.

In contrast, tensor network methods for three-dimensional (3D) statistical
models remain less explored. The primary computational bottleneck arises
from the high cost of contracting 3D network. For example, in the
gradient-based variational scheme, the partition function problem is
reformulated as a leading eigenvalue problem for a transfer matrix expressed
in the form of a projected entangled-pair operator (PEPO) and its leading
eigenvector represented by a projected entangled-pair state (PEPS). The
optimization of PEPS and the calculation of physical quantities require
contracting a triple-layer tensor network of PEPS and PEPO with a bond
dimension of $\mathcal{O}(D^{2}d)$ ($D$ and $d$ represent the bond
dimensions of PEPS and PEPO, respectively), which is computationally
expensive. Earlier attempts to use variational methods also encountered
problems related to numerical instability during optimization \cite%
{nishino_2000, nishino_2001, okunishi_2000, maeshima_2001a}. To the best of
our knowledge, variational approaches for 3D partition functions have been
limited to small bond dimensions ($D=2-4$) \cite{vanderstraeten_2018,
vanhecke_2022}. The generalization of other methods to three dimensions also
encounters challenges. For instance, a direct generalization of the CTMRG
method suffers from the extremely high order of bond dimension in
computational complexity and the slow decay of the spectrum of the density
matrices constructed by corner-transfer tensors \cite{nishino_1998,
orus_2012}. Additionally, extending the tensor renormalization group (TRG)
method to three dimensions faces challenging due to the increase in
computational complexity with bond dimension and the change in lattice
topology during renormalization \cite{xie_2012}.

To address the challenge of contracting tensor network in high dimensions,
various methods have been proposed to reduce the computational cost \cite%
{xie_2012, adachi_2020, zhang_2025}. One promising method is the nesting
technique, initially introduced in the context of calculating physical
quantities in 2D quantum systems \cite{xie_2017} and later adapted for
calculating physical quantities in 3D classical systems \cite{yang_2023}.
This technique compresses the double-layer tensor network into a
single-layer structure, which can then be contracted using the standard
multisite CTMRG algorithm. Recently, this method was further adapted for
gradient-based variational infinite PEPS (iPEPS) optimization in 2D quantum
systems using a carefully designed split-CTMRG scheme \cite{lan_2023,
naumann_2025}.

In this work, we extend the nesting idea and split-CTMRG method to
variational iPEPS optimization for 3D classical statistical models. The
crucial point for applying the nesting idea to variational method is to
contract the nesting network with high accuracy to evaluate the gradient.
Since the standard multi-site CTMRG method may not meet this requirement,
designing a scheme tailored for the nesting structure is needed. For 2D
quantum systems with local Hamiltonians, this central task has been
accomplished for double-layer networks \cite{naumann2024,naumann_2025}. To
tackle the 3D classical case which requires contracting both double- and
triple-layer networks, we propose a split corner transfer matrix (split-CTM)
environment for triple-layer network and an accompanying renormalization
group scheme. Our approach surpasses traditional methods in efficiency,
enabling variational optimization of iPEPS with larger bond dimensions.

The paper is organized as follows. In Sec.~\ref{section::gradient
optimization}, we show the basic idea of variational iPEPS optimization of a
3D classical model and the computation bottleneck for the traditional method. In
Sec.~\ref{section::split-CTMRG}, the triple-layer split-CTMRG scheme is
explained in detail. In Sec.~\ref{section::Benchmark: 3D classical Ising
model}, we take the 3D Ising model as a benchmark to show the performance of
this scheme. Finally, in Sec.~\ref{section:: Discussion and outlook},
we discuss potential improvements and generalizations of the algorithm.

\section{Variational iPEPS method for 3D classical statistical model}

\label{section::gradient optimization}

We begin by reviewing the gradient-based iPEPS optimization method \cite%
{corboz_2016,vanderstraeten_2018,vanderstraeten_2016}. To implement the
tensor network method, the first step is to represent the partition function
of the statistical model as a tensor network. As an example, consider
classical spin models with nearest-neighbor interactions on a cubic lattice
with the Hamiltonian given by
\begin{equation}
H=\sum_{\langle i,j\rangle }h(s_{i},s_{j}),
\end{equation}%
where $\langle i,j\rangle $ refers to nearest neighbors and $s_{i}$ denotes
the spin variables. The partition function can be decomposed into a tensor
network, expressed as the product of local Boltzmann weights on the lattice
links
\begin{equation}
\mathbb{Z}=\sum_{\{s_{i}\}}\prod_{<i,j>}e^{-\beta
h(s_{i},s_{j})}=\sum_{\{s_{i}\}}\prod_{<i,j>}W(s_{i},s_{j}).
\end{equation}%
This equation can be transformed into a tensor network, as shown in Fig.~\ref%
{Fig:tensor network representation of 3D statistical model}(a), where the $%
\delta $ tensor on the lattice sites is defined by
\begin{equation}
\delta (s_{1},s_{2},s_{3},s_{4},s_{5},s_{6})=%
\begin{cases}
1, & s_{1}=s_{2}=s_{3}=s_{4}=s_{5}=s_{6} \\
0, & \text{other cases.}%
\end{cases}%
\end{equation}%
This ensures that all indices of $W$ take the same value at the joint point.

\begin{figure}[t]
\centering
\includegraphics[width=1\linewidth]{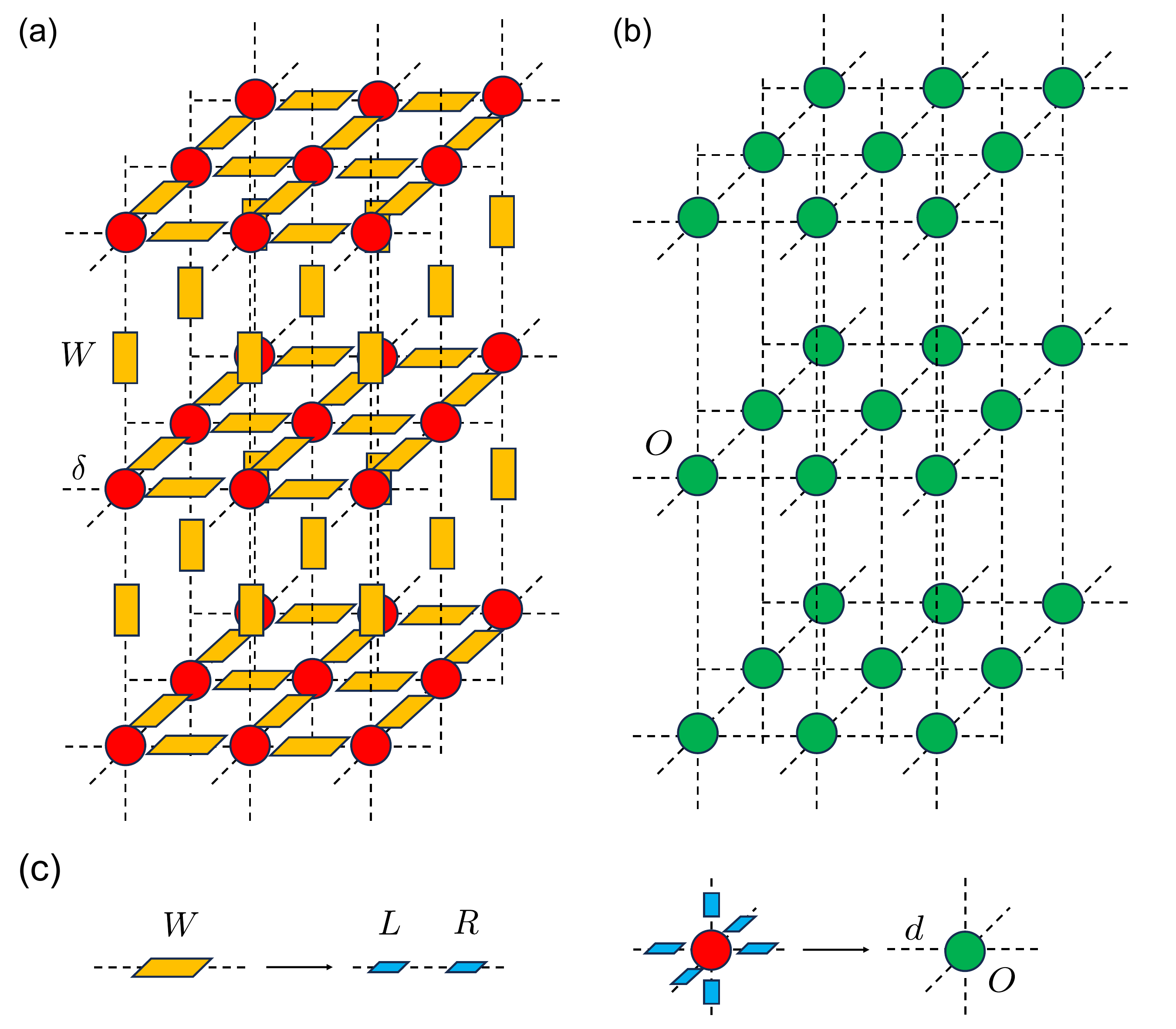}
\caption{Tensor network representation of the partition function for the 3D
classical model. (a) The tensor network (TN) representation, consisting of a
$\protect\delta$ tensor on each site and a $W$ matrix on each link. (b) The
TN representation consisting of a uniform tensor $O$. (c) The tensor $O$ is
built by grouping the $\protect\delta$ tensor with the connecting square
root of the $W$ matrix. The square root of the $W$ tensor is obtained via
singular value decomposition (SVD). The bond dimension of tensor $O$ is
labeled as $d$.}
\label{Fig:tensor network representation of 3D statistical model}
\end{figure}

To further transform the tensor network into a uniform one, we split the
matrix $W$ by using singular value decomposition
\begin{eqnarray}
W_{l}(s_{i},s_{j}) &=&\sum_{s_{k}}U(s_i,s_k)S(s_k)V(s_k,s_j)  \notag \\
&=&\sum_{s_{k}}\left( U\sqrt{S}\right) (s_{i},s_{k})\left( \sqrt{S}V\right)
(s_{k},s_{j})  \notag \\
&=&\sum_{s_{k}}L(s_{i},s_{k})R(s_{k},s_{j}).  \label{eqn: SVD}
\end{eqnarray}%
By grouping each $\delta $ tensor with the surrounding six $L$ (or $R$)
matrices into a single tensor $O$, the partition function can be represented
as a tensor network of uniform tensor $O$, as shown in Figs.~\ref{Fig:tensor
network representation of 3D statistical model}(b) and (c)
\begin{equation}
\mathbb{Z}=\text{tTr}\prod_{i}O_{i}(s_{1},s_{2},s_{3},s_{4},s_{5},s_{6}),
\label{eqn: TN for partition function}
\end{equation}
where ``$\text{tTr}$'' denotes the contraction over all auxiliary links in
the infinite 3D tensor network. The partition function can further be
reformulated as
\begin{equation}
\mathbb{Z}=\text{Tr}\{\hat{T}(O)^{N_{z}}\},
\end{equation}%
where $\hat{T}(O)$ is the plane-to-plane transfer matrix composed of a plane
of uniform local tensors $O$, and $N_{z}$ is the number of sites in the z direction of the system.

\begin{figure}[t]
\centering
\includegraphics[width=1\linewidth]{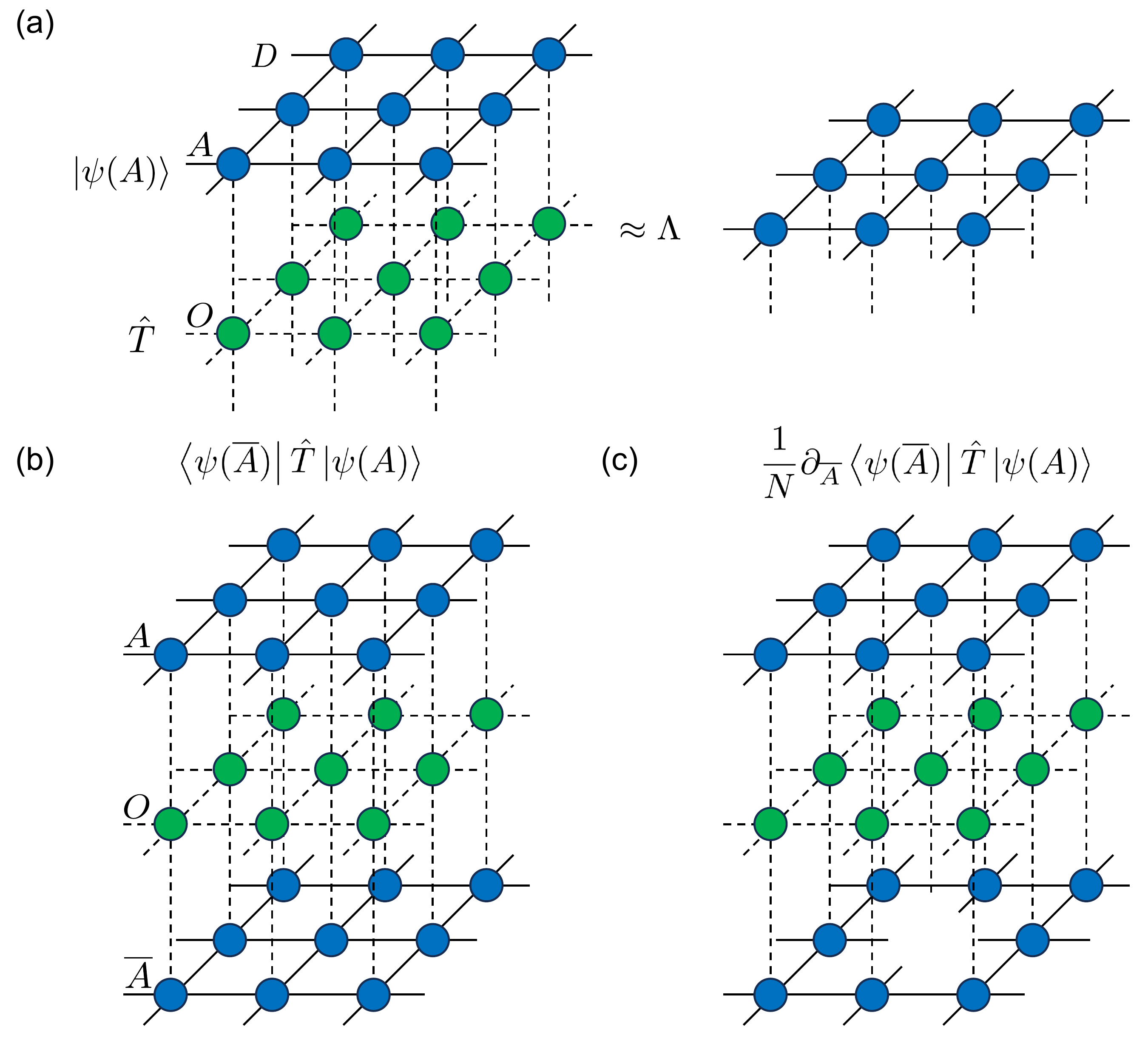}
\caption{iPEPS ansatz and the terms for calculating the free energy and
gradient. (a) The fixed point equation for the transfer matrix. The iPEPS
state $\ket{\psi(A)}$ and the plane-to-plane transfer matrix $\hat{T}$ are
made up of the uniform tensors $A$ and $O$, respectively. (b),(c)
Triple-layer network representations involved in the
calculation of free energy and its gradient. }
\label{Fig::iPEPS anstaz as fixed point}
\end{figure}

In the thermodynamic limit, the central task is to solve for the leading
eigenvalue and eigenvectors of the transfer matrix
\begin{equation}
\hat{T}(O)\ket{\psi}=\Lambda \ket{\psi},  \label{eqn: fixed-point equation}
\end{equation}%
where the leading eigenvector $\ket{\psi}$ can be approximated by using an
iPEPS, represented by a uniform tensor $A$ with virtual bond dimension $D$.
Equation ~\eqref{eqn: fixed-point equation} can then be represented
diagrammatically, as displayed in Fig.~\ref{Fig::iPEPS anstaz as fixed point}%
(a).

In the gradient-based optimization scheme, the eigen equation Eq.~%
\eqref{eqn: fixed-point equation} is transformed into an optimization
problem, aiming to minimize the cost function $f$ with respect to $A$,
\begin{equation}
f(A,\overline{A})=-\frac{1}{N\beta }\ln (\Lambda )=-\frac{1}{N\beta }\ln (%
\frac{\bra{\psi(\overline{A})}\hat{T}\ket{\psi(A)}}{\bra{\psi(\overline{A})}%
\ket{\psi(A)}}),  \label{eqn: free energy expression}
\end{equation}%
where $N=N_{x}N_{y}$ is the number of sites in the $xy$ plane of the system
and $f$ has the physical meaning of the free-energy density.

The gradient of the cost function is given by
\begin{eqnarray}
      g&&=2\times \partial _{\overline{A}}f(A,\overline{A})
\label{eqn: gradient expression} \nonumber\\
&&=-\frac{2}{N\beta }\left( \frac{\partial _{\overline{A}}%
\bra{\psi(\overline{A})}\hat{T}\ket{\psi(A)}}{\bra{\psi(\overline{A})}\hat{T}%
\ket{\psi(A)}}-\frac{\partial _{\overline{A}}\bra{\psi(\overline{A})}%
\ket{\psi(A)}}{\bra{\psi(\overline{A})}\ket{\psi(A)}}\right) .  
\end{eqnarray}
Since both $\hat{T}$ and $\ket{\psi(A)}$ are represented as tensor networks,
the terms in the free energy density and the gradient can be represented as
infinite 2D network, as shown in Figs.~\ref{Fig::iPEPS anstaz as fixed
point}(b) and (c). By using an approximate tensor network contraction method, one
can compute the gradient and the value of the function with respect to $A$.
Various optimization methods, such as the quasi-Newton method or
conjugate-gradient method, can then be applied to iteratively solve for $A$.

\begin{figure}[t]
\centering
\includegraphics[width=1\linewidth]{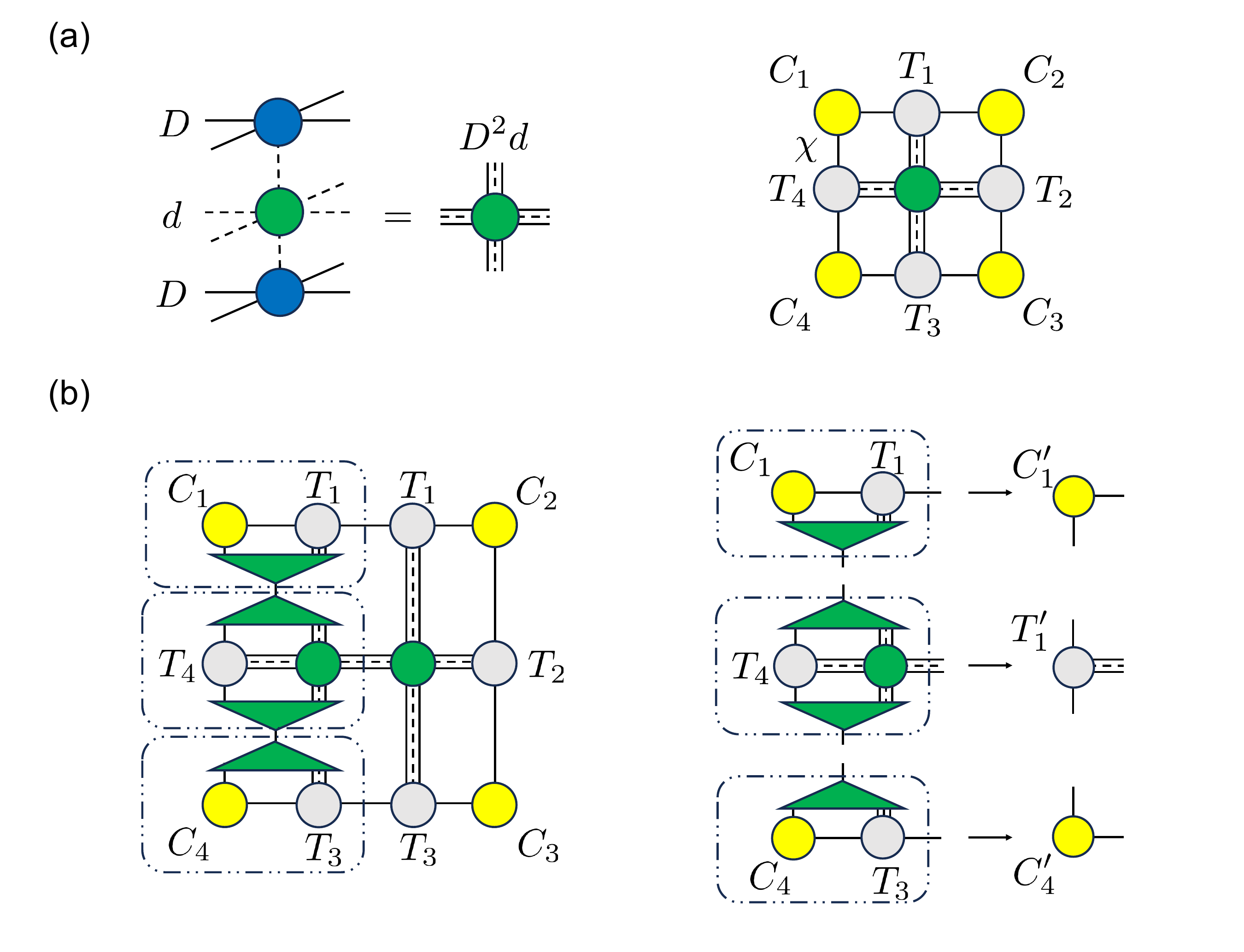}
\caption{Reduced CTMRG scheme for contracting the triple-layer network. (a)
The environment tensor configuration for $C$ and $T$. (b) The absorption and
truncation procedure during a left move.}
\label{Fig::reduced method review}
\end{figure}

A key bottleneck in this algorithm is the evaluation of the free energy and
its gradient, which involves contracting a triple-layer infinite tensor
network at each optimization step. In the traditional reduced method, the
contraction is first performed by contracting the physical indices
connecting the different layers, resulting in a single-layer network, as
shown in Fig.~\ref{Fig::reduced method review}(a). The network can then be
contracted by applying standard 2D tensor network contraction techniques.
One commonly used technique is the CTMRG with directional moves \cite%
{orus_2009,corboz_2011,corboz_2014}, as illustrated in Fig.~\ref%
{Fig::reduced method review}(b). In this approach, environment tensors,
including corner tensors $\{C\}$ and edge tensors $\{T\}$ with boundary bond
dimension $\chi$, are used to approximate the environment after contracting
the infinite network. The $\{C\}$ and $\{T\}$ tensors are solved using the
power method. In each step, the system grows in one direction, and the local
tensors from the system are absorbed into the environment tensors, which are
subsequently truncated using projectors $P$.

\begin{figure*}[t]
\centering
\includegraphics[width=1\linewidth]{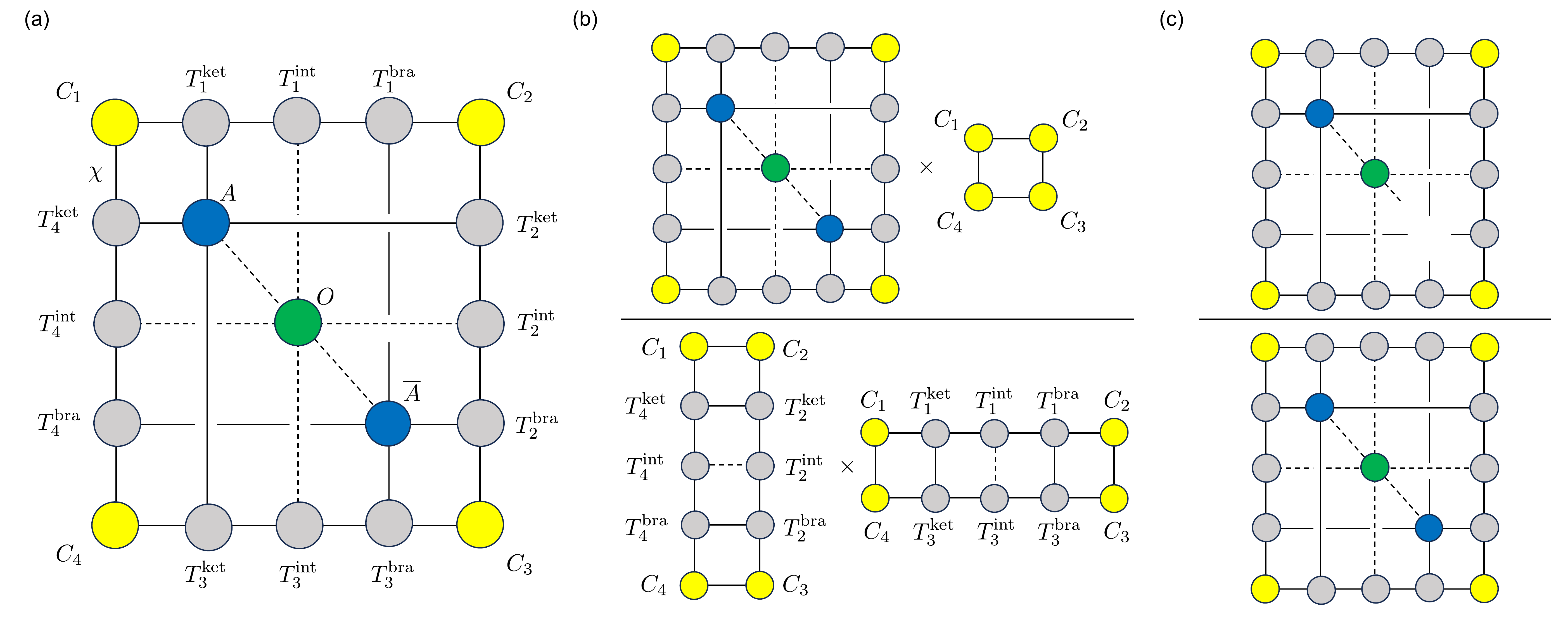}
\caption{The split-CTM environment tensor setting. The bonds of the $O$ in PEPO
are denoted by dashed lines, while the bonds of the PEPS and the environment
are denoted by solid lines. (a) The split-CTM environment tensor
configuration for the triple-layer network. (b) The numerator involved
in calculating the free energy. (c) The term involved in calculating the
gradient.}
\label{Fig::Split method tensor setting}
\end{figure*}

For the triple-layer structure contracted by reduced method, the leading
computation time of contraction is in the scale of \footnote{%
The complexity analysis in this article is taken under the assumptions that $%
d$ are relatively small compared with $D$ and empirically $\chi $ should be
at least the order of $D^{2}$.}
\begin{equation}
\mathcal{O}(\chi ^{3}D^{4}d^{3})\sim \mathcal{O}(D^{10}d^{3})
\end{equation}
under the help of a partial singular value decomposition (SVD) technique aiming at solving the largest $\chi$
singular values with iterative solver \cite{motoyama2022}. The leading
memory space cost is
\begin{equation}
\mathcal{O}(\chi ^{2}D^{4}d^{2})\sim \mathcal{O}(D^{8}d^{2}).
\end{equation}

The high computation cost restricts the accessible bond dimension of iPEPS
tensors in variational method. Previous work can only get reliable results
with $D\leq 4$ for 3D classical Ising model \cite%
{vanderstraeten_2018,vanhecke_2022}.

\section{Triple-layer split-CTMRG scheme}

\label{section::split-CTMRG}

To reduce the computational cost associated with contracting a triple-layer
tensor network, we propose a split-CTMRG method. Compared to the
traditional reduced method, the split-CTMRG introduces two key differences.
First, the local tensors with a triple-layer structure are compressed into a
single-layer tensor network, where the physical indices are not contracted.
Correspondingly, the environment edge tensors are decomposed into separate
edge tensors for each layer. Second, while the local tensors corresponding
to the original triple layers are absorbed in one move, the projectors are
constructed and applied to the environment tensors in a separate, sequential
manner. The details of the split-CTMRG method are presented below.

\subsection{Environment tensor setting}

\label{subsection::Environment tensor setting}

As shown in Fig.~\ref{Fig::Split method tensor setting}(a), for a
triple-layer tensor network, the environment tensors in the split-CTMRG
scheme are set as $E=\{C_{i},T_{i}^{\text{ket}},T_{i}^{\text{int}},T_{i}^{%
\text{bra}},i=1,2,3,4\}$, where $T^{\text{ket}}$, $T^{\text{int}}$, and $T^{%
\text{bra}}$ represent the edge tensors for each layer, and $%
\{C_{i},i=1,2,3,4\}$ are the four corner matrices. In contrast to the
original nested tensor network approach, which introduces swap tensors to
transform the single-layer network into a regular multisite network with
regular lattice geometry (which is then contracted via a multisite CTMRG
scheme)\cite{xie_2017}, the split-CTMRG scheme directly projects the
triple-layer network into a single-layer representation. In this
representation, the local tensors $O$ of the partition function are directly
connected to the internal edge environment $T^{\text{int}}$. Furthermore,
the structure of the environment tensors differs from that of the multisite
CTMRG method, where the number of corner, edge tensors scale as $4n_{x}n_{y}
$ and $n_{x}\times n_{y}$ being the unit cell size. For a double-layer
tensor network, the split-CTM environment is constructed in a similar manner
as for the triple-layer case, with four corner tensors and four edge tensors
per layer, labeled as $T^{\text{ket}}$ and $T^{\text{bra}}$.

Under this environment setup, both the free energy and gradient terms can be
expressed in terms of the CTM tensors, as illustrated in Figs.~\ref%
{Fig::Split method tensor setting}(b),(c). Here, we focus on the contraction
of the triple-layer network and ignore the terms related to the
normalization of $|\psi (A)\rangle $.

\subsection{Renormalization scheme}

\label{subsection::directional move}
\begin{figure*}[t]
\centering
\includegraphics[width=1\linewidth]{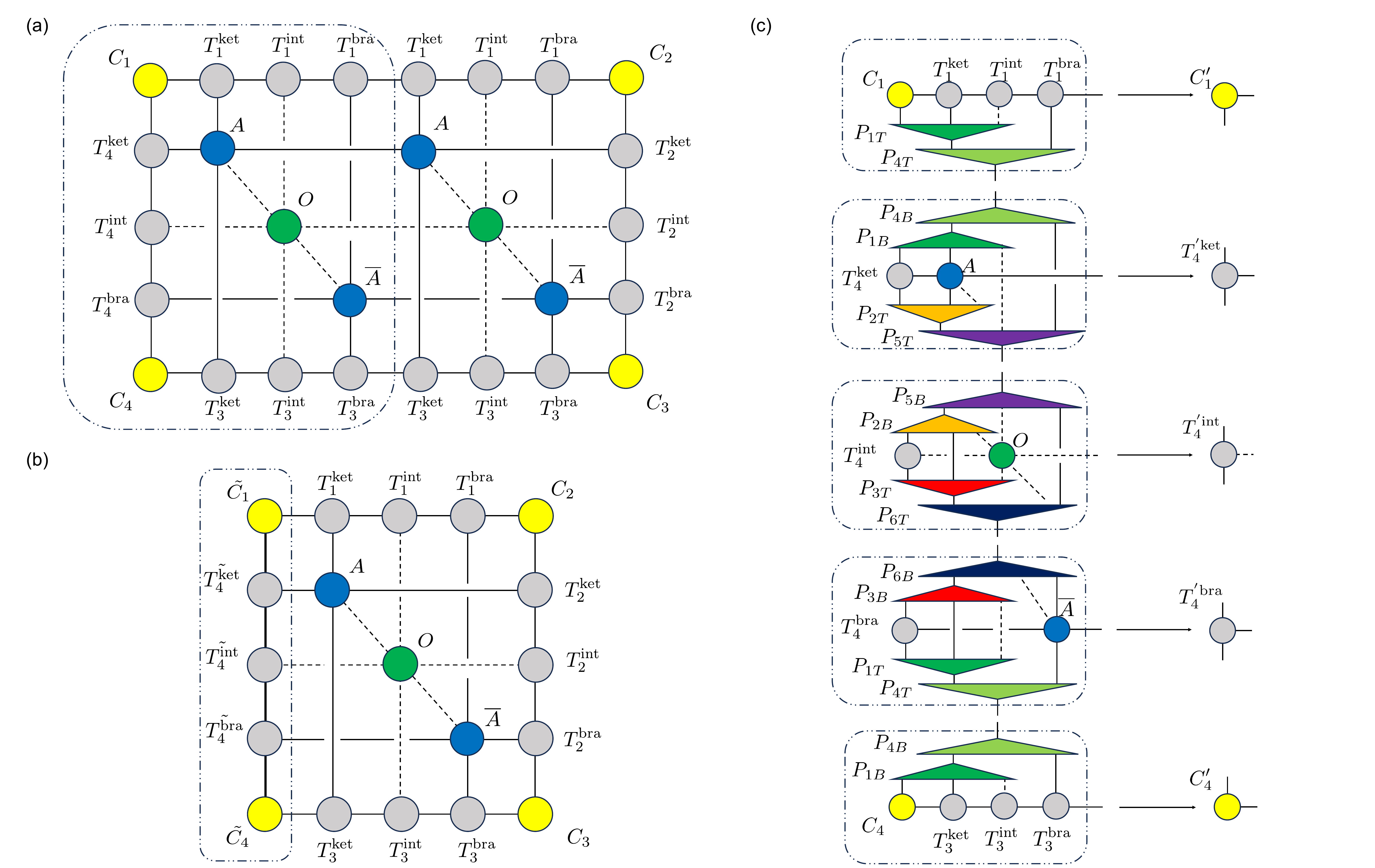}
\caption{Left move in the split-CTMRG scheme for the triple-layer network.
(a) Insertion. (b) Absorption. (c) Truncation.}
\label{Fig::RG scheme for triple-layer split-CTMRG}
\end{figure*}
For a general PEPO without any specific symmetry assumptions, the
renormalization of the environment tensors is performed using directional
moves. A complete split-CTMRG update consists of four directional moves:
left, right, up, and down, which are applied sequentially until the
environment converges. Here we take the left move as an example to
illustrate the CTMRG procedure, as shown in Fig.~\ref{Fig::RG scheme for
triple-layer split-CTMRG}.

The main steps of the procedure are as follows:

\begin{enumerate}
\item As shown in Fig.~\ref{Fig::RG scheme for triple-layer split-CTMRG}%
(a), insert a new set of tensors $\{T_{1}^{\text{ket}},T_{1}^{\text{int}%
},T_{1}^{\text{bra}},T_{3}^{\text{ket}},T_{3}^{\text{int}},T_{3}^{\text{bra}%
},A,O,\overline{A}\}$.

\item The newly inserted tensors in each row are contracted into the left
environment tensors $\tilde{C}_{1}$, $\tilde{T}_{4}^{\text{ket}}$, $\tilde{T}%
_{4}^{\text{int}}$, $\tilde{T}_{4}^{\text{bra}}$, and $\tilde{C}_{4}$ [see
Fig.~\ref{Fig::RG scheme for triple-layer split-CTMRG}(b)], resulting in the
increase of vertical bond dimension. 

\item In the split-CTM scheme, the isometric projectors are decomposed into
multiple smaller projectors to reduce computational cost. Both the form and
construction of the projectors have a certain degree of arbitrariness. The
choice of whether to absorb and truncate the intermediate layer
independently or in conjunction with the ket and bra layers leads to
different schemes. Here, we present a six-projector scheme in which the
bonds of the PEPO tensor $O$ are truncated alongside the projectors for the
ket and bra layers. An alternative nine-projector scheme, in which the
intermediate physical PEPO is absorbed and truncated independently, is
discussed in Sec.~\ref{section:: Discussion and outlook}.
\end{enumerate}

As illustrated in Fig.~\ref{Fig::RG scheme for triple-layer split-CTMRG}(c),
isometric projectors $\{P_{aT},P_{aB},a=1,2,3,4,5,6\}$ are inserted along
the vertical bonds to reduce the bond dimension. The isometry projectors
with same index numbers are in the same group, satisfying the relation: $%
P_{aT}P_{aB}=I$. For each vertical bond to be truncated, two projectors are
used: one to truncate the bonds related with the ket layer and the other to
truncate the bond related to the bra layer, This yields the updated
environment tensors $C^{\prime }_1$, $T_4^{^{\prime }\text{ket}}$, $%
T_4^{^{\prime }\text{int}}$, $T_4^{^{\prime }\text{bra}}$, and $C^{\prime
}_4 $. The above procedure can be generalized to other directions
straightforwardly.

In certain cases, lattice symmetries are preserved at the level of the
transfer matrix and the local tensors. Exploiting these symmetries can
further reduce the computational cost. For instance, in a cubic lattice, if
the transfer matrix is real and symmetric, and the local PEPO tensor $O$ is
invariant under the $O_h$ symmetry group, the PEPS tensor $A$ can be chosen
to be real, with its virtual bonds constrained to be invariant under the $D_4
$ symmetry group. Symmetry constraints can be imposed at each optimization
step by projecting both the tensor $A$ and its gradient $g$ onto the
symmetric subspace.

\begin{figure}[b]
\centering
\includegraphics[width=\linewidth]{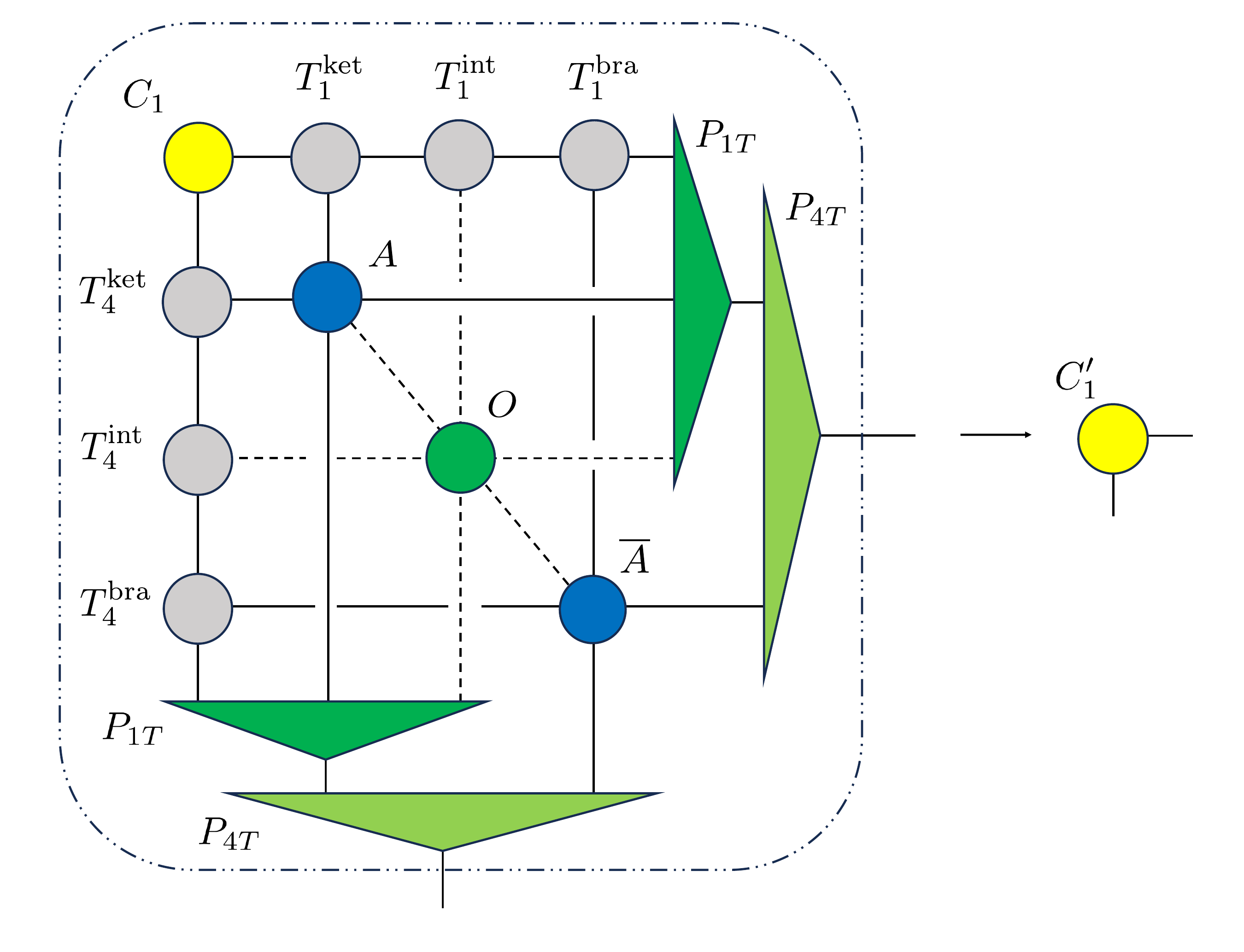}
\caption{Symmetric absorption and truncation of corner tensors in the
triple-layer split-CTMRG scheme. }
\label{Fig::D2 symmetric CTMRG}
\end{figure}

Moreover, in the split-CTMRG scheme, the nested structure reduces the
symmetry from $D_{4}$ to $D_{2}$, leading to the following constraints on
the environment tensors:
\begin{equation}
\begin{aligned} C_1&=C_3,C_2=C_4\\ C_i&=C_i^T,\quad i=1,2,3,4\\
T_4^{\text{ket}}&=(T_1^{\text{ket}})^T=T_2^{\text{bra}}=(T_3^{\text{bra}})^T%
\\
T_4^{\text{bra}}&=(T_1^{\text{bra}})^T=T_2^{\text{ket}}=(T_3^{\text{ket}})^T%
\\
T_4^{\text{int}}&=(T_1^{\text{int}})^T=T_2^{\text{int}}=(T_3^{\text{int}})^T.
\end{aligned}  \label{Symmetry constraints}
\end{equation}%

\begin{figure*}[t]
\centering
\includegraphics[width=1\linewidth]{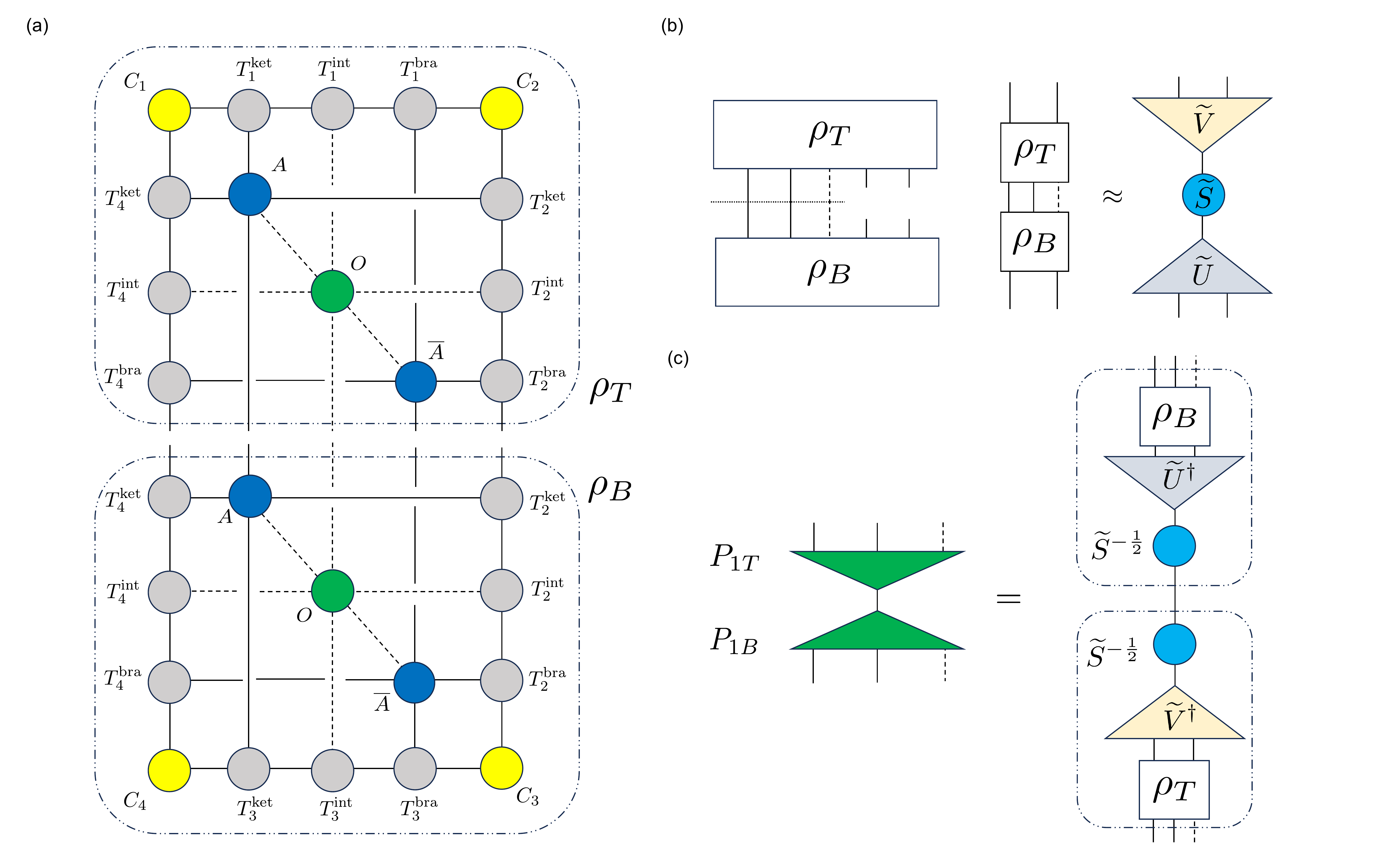}
\caption{Construction of projector 1. (a)~$\protect\rho_T$ and $\protect\rho%
_B$ consist of patches of tensors representing the top and bottom
environments. (b)~SVD is performed in the $\protect\rho_B\protect\rho_T$,
yielding $\widetilde{U},\widetilde{S},\widetilde{V}$. The largest $\protect%
\chi$ singular value is kept. (c)~A pair of projectors are constructed
to approximate isometry.}
\label{Fig::construction of projector 1}
\end{figure*}
Here $T_i^T$ means the transpose operation on the virtual bonds of edge
tensors. With these symmetry constraints, a symmetric split-CTMRG scheme
becomes more efficient than the directional move scheme, since all corner
and edge tensors can be updated using a single set of projectors. While the
edge tensors are renormalized in the same manner as in the directional
scheme, the absorption and truncation of corner tensors is done in a
different way, as illustrated in Fig.~\ref{Fig::D2 symmetric CTMRG}.

\subsection{Projectors construction}
\label{subsection::Projectors construction}

Projectors serve as approximate isometries inserted between the bonds to be
truncated. The construction scheme is generally not unique. Here we apply
the commonly used method developed in Ref.~\cite{corboz_2014}.

As an example, we first illustrate the construction of projectors $P_{1T}$
and $P_{1B}$. As depicted in Fig.~\ref{Fig::construction of projector 1}(a),
we consider an effective environment consisting of $1\times 2$ unit cells of
the local tensor set $\{A,O,\overline{A}\}$ along with the surrounding
environment tensors to build the projector. The environment is further
divided into two parts: one corresponding to the upper-half environment $%
\rho _{T}$ and the other to the lower-half environment $\rho _{B}$. The
dangling indices of $\rho _{T}$ and $\rho _{B}$ are then regrouped into two
sets: one set contains the indices to be truncated, while the other index
contains the remaining indices. Thus, $\rho _{T}$ and $\rho _{B}$ can be
treated as matrix.

After inserting a pair of projectors, the matrix $\rho_B P_{1B} P_{1T} \rho_T
$ will serve as a low-rank approximation of $\rho _{B}\rho _{T}$. The
optimal approximation can be achieved by performing singular value
decomposition to $\rho _{B}\rho _{T}$, as shown in Fig.~\ref%
{Fig::construction of projector 1}(b),
\begin{equation}
\rho _{B}\rho _{T}=USV,
\end{equation}%
and set
\begin{equation}
\rho_B P_B=\widetilde{U}\widetilde{S}^{\frac{1}{2}},\quad P_T \rho_T=%
\widetilde{S}^{\frac{1}{2}}\widetilde{V},
\label{eqn: projector construction}
\end{equation}%
where $\widetilde{S}$ is the truncated singular spectrum and $\widetilde{U},%
\widetilde{V}$ are the corresponding truncated singular vectors. To satisfy
Eq.~\eqref{eqn: projector construction}, the projectors can be constructed
as shown in Fig.~\ref{Fig::construction of projector 1}(c),
\begin{equation}
P_{1B}=\rho _{T}\widetilde{V}^{\dagger }(\widetilde{S}^{\frac{1}{2}%
})^+,\quad P_{1T}=(\widetilde{S}^{\frac{1}{2}})^+\widetilde{U}^{\dagger
}\rho _{B},  \label{eqn: projector construction 2}
\end{equation}
where $(\widetilde{S}^{\frac{1}{2}})^+$ is the pesudo inverse of $\widetilde{%
S}^{\frac{1}{2}}$.

The other five projectors used in Fig.~\ref{Fig::RG scheme for triple-layer
split-CTMRG} can be calculated in a similar manner to $P_{1T},P_{1B}$ but
with different choices of the effective environment $\rho _{T}$ and $\rho
_{B}$. While the detailed constructions of each projector are presented in
the Appendix, we briefly discuss the general principle for selecting $\rho
_{T}$ and $\rho _{B}$.

There is a trade-off between efficiency and accuracy when choosing $\rho
_{T} $ and $\rho _{B}$. While using larger tensors for $\rho _{T}$ and $\rho
_{B}$ may improve accuracy by capturing a more relevant basis, it can also
increase the computational cost. We observe that $\rho _{B}\rho _{T}$ can be
viewed as a successive product of four matrices at the corner (connected by
the virtual bonds) and the number of non zero singular values is bounded by
the smallest rank of these matrices. Hence, we also make sure that the bond
dimensions of the four corners are larger than the bond dimensions to be
truncated.

The leading computation cost of the entire scheme arises during the
construction of projector 4, which can be reduced to a scale of
\begin{equation}
\mathcal{O}(\chi ^{3}D^{3}d^{3})\sim \mathcal{O}(D^{9}d^{3}),
\end{equation}%
while the memory cost is reduced to
\begin{equation}
\mathcal{O}(\chi ^{2}D^{3}d^{2})\sim \mathcal{O}(D^{7}d^{2}).
\end{equation}

\subsection{Convergence criterion}

The singular value spectrum of the effective density matrix formed by the
four corner tensors is used to check the convergence of the algorithm. The
effective density matrix $\rho _{\text{eff}}$ is given by:
\begin{equation}
\rho _{\text{eff}}=\frac{C_{1}C_{2}C_{3}C_{4}}{\text{Tr}%
(C_{1}C_{2}C_{3}C_{4})}=USV^{T},
\end{equation}%
where $\rho _{\text{eff}}$ represents the effective density matrix of the
system, and $S$ is the diagonalized matrix containing the singular-values
from the decomposition.

Convergence is determined by monitoring the change in the singular value
spectrum between two successive steps. If the difference between the spectra
is below a specified threshold, the CTMRG procedure is considered to have
converged
\begin{equation}
\Vert S^{[i+1]}-S^{[i]}\Vert _{2}<\delta ,
\end{equation}%
where $||\cdot ||_{2}$ denotes the $l_{2}$ norm.

\section{Benchmark: 3D Ising model}

\label{section::Benchmark: 3D classical Ising model}

We apply our method to the 3D classical Ising model on a cubic lattice as a
benchmark. The Hamiltonian of the Ising model in the absence of an external
field is given by
\begin{equation}
H=-\sum_{\langle i,j\rangle}\sigma_i\sigma_j,
\end{equation}
where $\sigma_i=\pm 1$. The Boltzmann weight on each link is represented as
\begin{equation}
W=%
\begin{pmatrix}
e^{\beta } & e^{-\beta } \\
e^{-\beta } & e^{\beta }%
\end{pmatrix}%
,
\end{equation}%
where $\beta $ is the inverse temperature. Since $W$ is a positive-definite,
real symmetric matrix, its decomposition can be performed via eigenvalue
decomposition rather than the SVD described in Eq.~\eqref{eqn: SVD}.
Specifically, we write
\begin{equation}
W = U \lambda U^{-1} = (U \lambda^{\frac{1}{2}} U^{-1})(U \lambda^{\frac{1}{2%
}} U^{-1}) = \sqrt{W} \sqrt{W},
\end{equation}
where $U$ is the orthogonal matrix of eigenvectors and $\lambda$ is the
diagonal matrix of eigenvalues. $\sqrt{W}$ denotes the matrix square root of
$W$.

By contracting the $\delta$ tensor with the surrounding $\sqrt{W}$ matrices,
the PEPO representation of the Ising model is constructed. Importantly, the
local PEPO tensor preserves the full cubic point group symmetry $O_h$.
Accordingly, we employ the symmetric split-CTMRG method for iPEPS
optimization of the 3D Ising model.

\subsection{Accuracy and efficiency}

Before discussing the physical results, we first demonstrate that the free
energy density and the gradient of the system can be calculated accurately,
with a significant speedup achieved using the split-CTMRG scheme. This
speedup without losing accuracy is crucial for enabling the use of larger
bond dimensions for tensor $A$ in gradient-based optimization.

\begin{figure}[t]
\centering
\includegraphics[width=\linewidth]{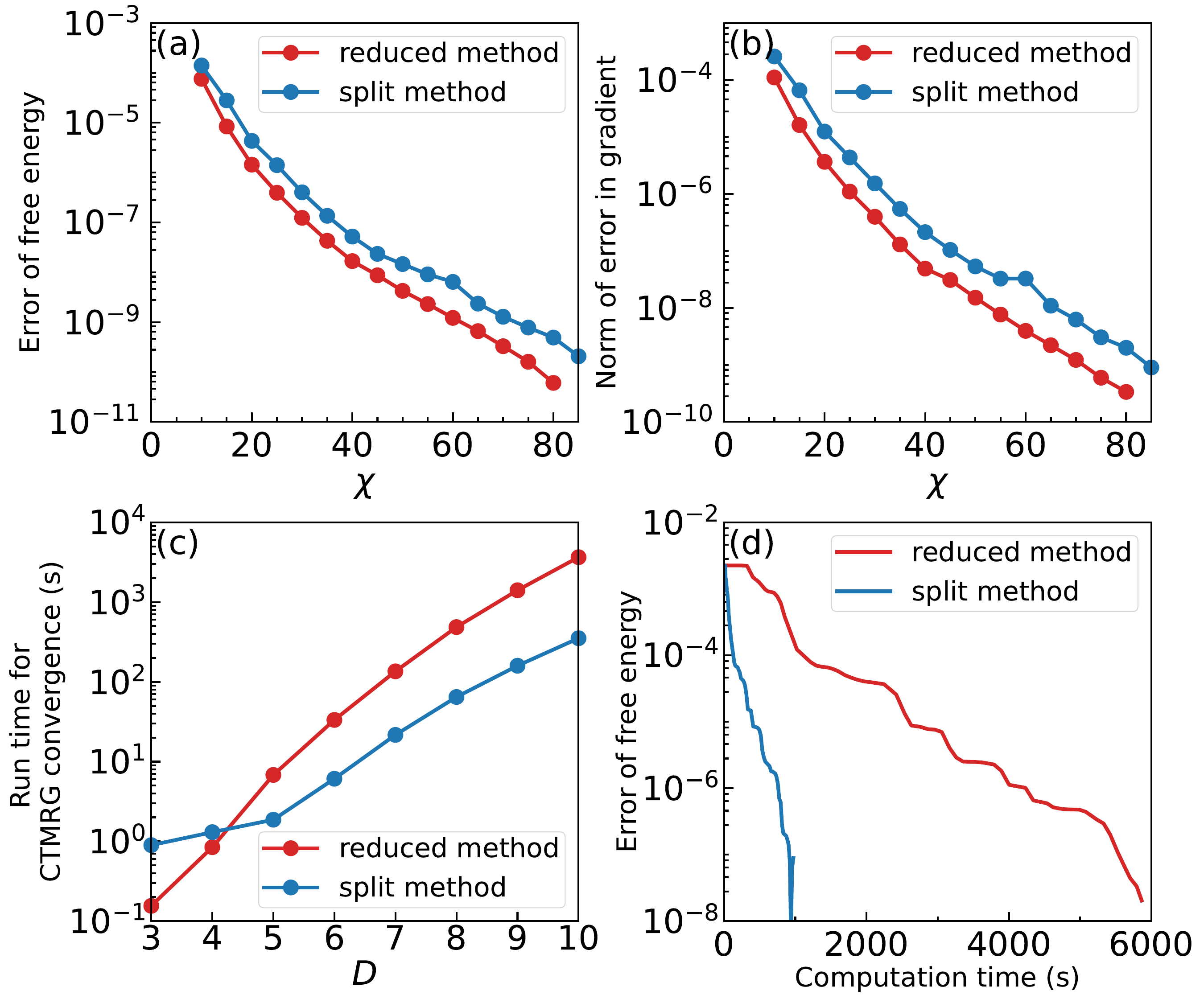}
\caption{Benchmark of accuracy and efficiency for the split-CTMRG scheme.
(a),(b) Comparison of the error in free energy and the norm of the error in
the gradient for two methods for different bond dimensions $\protect\chi$. Data
points are taken at $T=4.4$ with $D=6$. (c) Runtime for CTMRG convergence.
The contraction is performed with a converged $A$ at $T=4.4$ for various values 
$D$ with $\protect\chi =D^{2}$. The convergence criterion is taken as
$\protect\delta =10^{-10}$. (d) Error in free energy for the two methods as
a function of computation time at $T=4.4$ with $D=6$ and $\protect\chi=60$.
The error is calculated relative to the converged result of the reduced
method.}
\label{Fig::CTMRG benchmark}
\end{figure}

We contract the double-layer and triple-layer network for an Ising PEPO
tensor $O$ at a certain temperature and an optimized iPEPS tensor $A$, using
both the reduced and split CTMRG methods to calculate free energy and
gradient. Figures ~\ref{Fig::CTMRG benchmark}(a) and (b) present the error of
free energy $|f(\chi)-f_{\text{re}}(\chi_{\text{ref}})|$ and the norm of
error in gradient $||g(\chi)-g_{\text{re}}(\chi_{\text{ref}})||_\infty$
calculated with two methods respectively. The results from reduced method
with sufficiently large bond dimension $\chi_{\text{ref}}=85$ taken as a
reference. The comparison shows that the split method could yield results
comparable to the reduced method though a larger $\chi$ is required to reach
the same accuracy.

\begin{figure}[t]
\centering
\includegraphics[width=\linewidth]{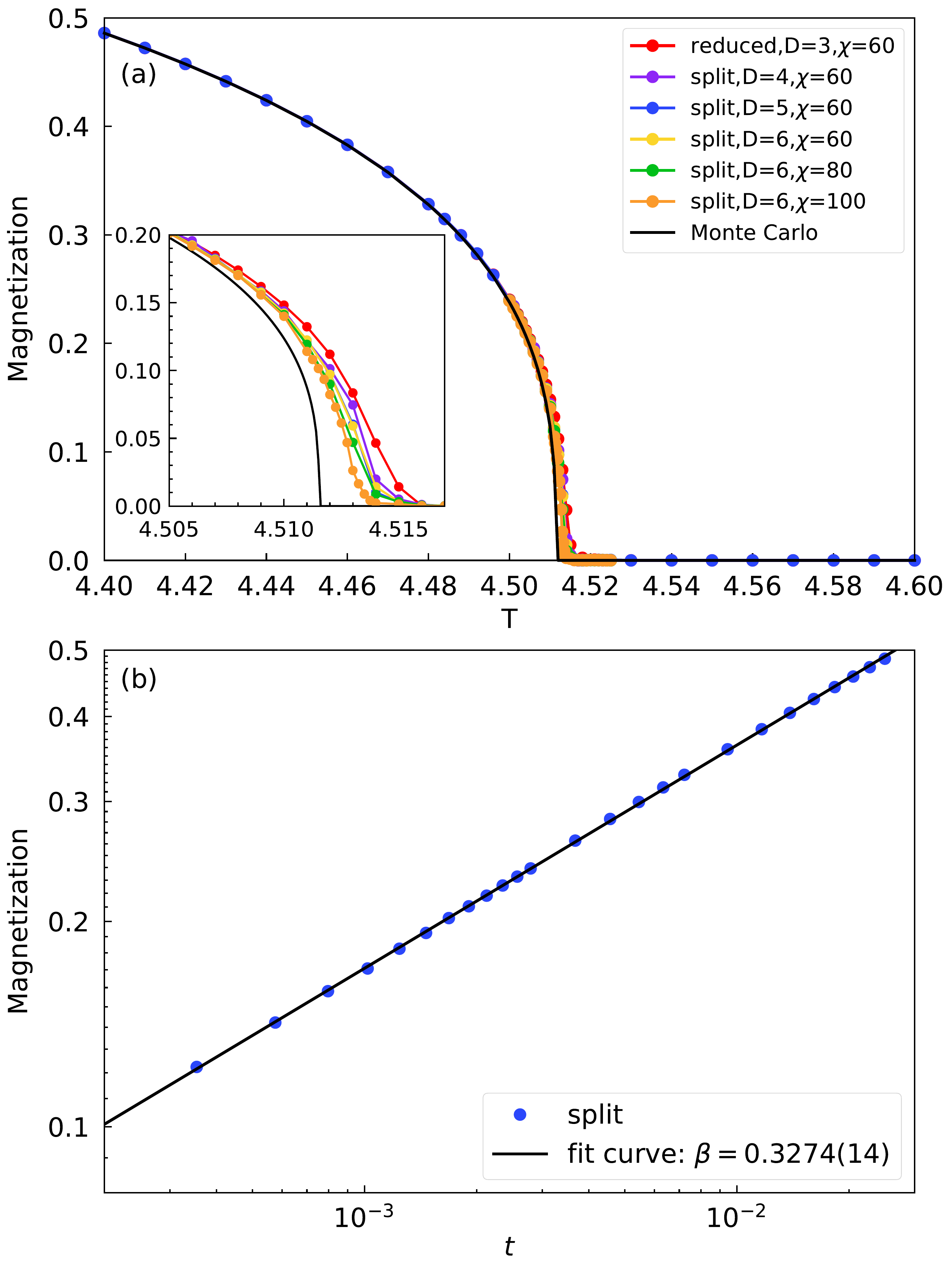}
\caption{(a) Magnetization for the 3D Ising model as a function of
temperature, computed using different methods and bond dimensions. Inset is
the magnetization curve in the critical region. (b) Fit of the critical exponent
$\protect\beta$. Magnetization data from $T=4.4$ to $T=4.511$ are used for
fitting.}
\label{Fig::Magnetization and fit}
\end{figure}

\begin{figure}[t]
\centering
\includegraphics[width=\linewidth]{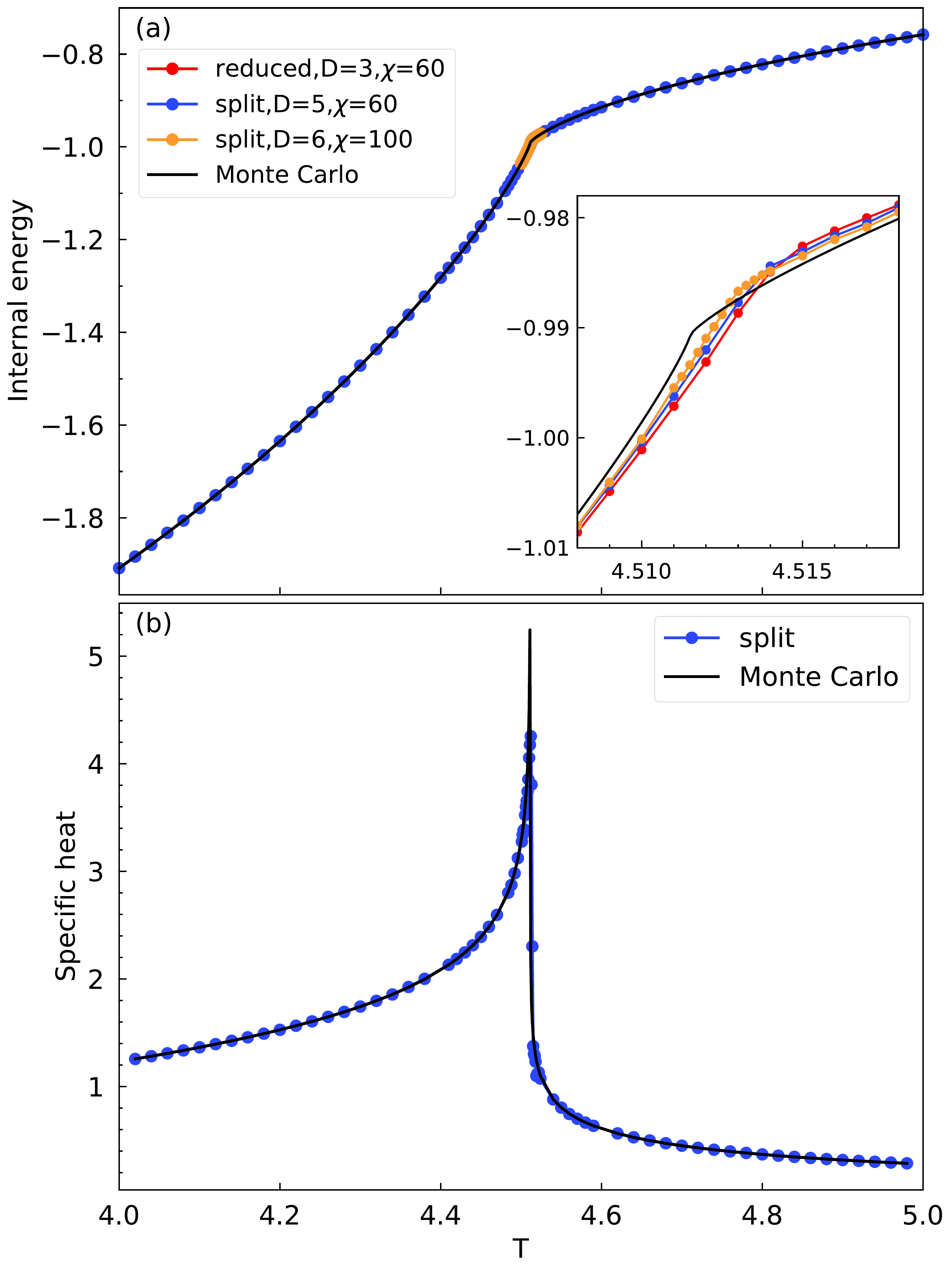}
\caption{(a) Internal energy and (b) specific heat for the 3D Ising model as
a function of temperature, computed using different methods and bond
dimension.}
\label{Fig::U and Cv}
\end{figure}
Compared to the reduced method, the split method has a lower leading
computation cost, though it comes with a larger prefactor that scales
linearly with the number of projectors. In our tests, the split method
becomes more efficient than the reduced method when the bond dimension
exceeds $D=5$, with the advantage becoming more pronounced as the bond
dimension increases, as shown in Fig.~\ref{Fig::CTMRG benchmark} (c).

When applied to gradient-based variational optimization, the split method
also shows an obvious speedup, as shown in Fig.~\ref{Fig::CTMRG benchmark}%
(d). Here we use the error of free energy $|f(t)-f_{\text{re}}(t_{\text{conv}%
})|$ with the converged result using reduced method as a reference to
present the convergence procedure of optimization. In the tested cases,
optimization using the split method requires approximately the same number
of steps to converge to the same accuracy as the reduced method. Therefore,
the speedup for optimization is similar to that for each step of CTMRG
convergence.

\subsection{Physical quantities}

This section presents the results of thermal quantities calculated using the
split-CTMRG scheme. The optimization of the iPEPS tensor $A$ was performed
via the limited-memory Broyden-Fletcher-Goldfarb-Shanno method, with the
convergence criterion set to $\Vert g\Vert _{\infty }<10^{-7}$. Physical
quantities were subsequently computed using the split-CTM environment, in
which the local Ising PEPO tensor is replaced by the corresponding impurity
tensor.

The temperature dependence of magnetization $M$ obtained using both the
reduced method and the split method, in comparison with the Monte Carlo
result\cite{talapov1996,feng_2010}, is shown in Fig.~\ref{Fig::Magnetization
and fit}(a). The results from the split method with same bond dimension ($%
D=3, \chi=60$) exhibit excellent agreement with those from the reduced
method. In the off-critical regions, the magnetization obtained through the
split method with different bond dimensions is consistent with the Monte
Carlo result. While in the critical region, as the bond dimension increases (%
$D=3 - 6, \chi=60 - 100$), the magnetization curve converge towards the
Monte Carlo results. From the singular point of the magnetization curve, the
critical temperature is estimated to be $T_{c}=4.51288(13)$, which is
comparable with the Monte Carlo result \cite{Hasenbusch_2010,
ferrenberg_2018} $T_{c}=4.51152326(11)$ and the high-order tensor
renormalization group (HOTRG) result \cite{wang2014} $T_c=4.51152469(1)$.

Furthermore, as shown in Fig.~\ref{Fig::Magnetization and fit}(b), by
fitting the numerical data of $M$ in the near-critical region with the
formula:
\begin{equation}
M\sim t^{\beta },
\end{equation}%
where $t=\frac{|T-T_{c}|}{T_{c}}$, we obtain the critical exponent $\beta
=0.3274(14)$, which is consistent with the previous Monte Carlo result $%
\beta =0.3263$ (for 3D Ising model \cite{ferrenberg_2018}), $\beta
=0.32642(4)$ for the model in the 3D Ising universality class \cite%
{Hasenbusch2021}, and the HOTRG result \cite{xie_2012} $\beta =0.3295$.

The temperature dependence of the internal energy $U$ and specific heat $C_v$
are shown in Fig.~\ref{Fig::U and Cv}, where $C_v$ is obtained by taking the
numerical derivative of the $U(T)$. These two quantities also agree well
with the Monte Carlo results over a wide temperature range, with slight
deviations near the critical point. From the peak of heat capacity, the
critical temperature is estimated as $T_c=4.5120(5)$, which is consistent
with the value obtained from the magnetization curve. These results indicate
that the split-CTMRG works well for the 3D classical model.

\section{Discussion and outlook}

\label{section:: Discussion and outlook}

In this paper, we introduce a split-CTMRG scheme designed to accelerate the
variational optimization of three-dimensional statistical models. The
proposed method achieves a reduced leading computational cost of $\mathcal{O}%
(D^{9})$, demonstrating a clear efficiency advantage over the traditional
reduced method for the 3D classical Ising model when $D\geq 5$. Using this
approach, we obtain iPEPS tensors with large bond dimensions (e.g., $D=6$),
and the physical results are in agreement with the state-of-the-art Monte
Carlo and the other schemes of tensor network calculations.

We compare our variational method with another powerful approach, the HOTRG
method for solving three-dimensional classical models. For HOTRG, the
computational cost scales as $\mathcal{O}(\widetilde{D}^{11})$, where $%
\widetilde{D}$ denotes the truncation bond dimension of the TRG tensors. In
contrast, the computational cost in the split-CTM scheme is controlled by
two parameters and scales as $\mathcal{O}(D^{3} \chi^{3})$, where $D$ is the
bond dimension of the iPEPS tensor and $\chi$ is the bond dimension of the
CTM environment. By applying the empirical relation $\chi \sim D^{2}$, the
computational cost can be estimated as $\mathcal{O}(D^{9})$.

However, during the numerical optimization in the variational method, a
large number of iterations (typically on the order of $\mathcal{O}(10^{2})$
steps) are required to achieve convergence, leading to a substantial
prefactor in the actual computational time. Consequently, although the
asymptotic scaling of our method is less than that of HOTRG, the maximum
bond dimension ($D=6$) achieved by our method is smaller than the
corresponding $\widetilde{D}=23$ reached by HOTRG \cite{wang2014}.
Nevertheless, the variational method offers the advantage of providing
relatively accurate results even for small bond dimensions. For instance,
our results at $D=4$ are already comparable to those obtained by HOTRG at $%
\widetilde{D}=12$ \cite{xie_2012}. While our current results $(D=6, \chi=100)
$ are not as accurate as the best HOTRG results $(\widetilde{D}=23)$. In
principle, our method is capable of handling larger bond dimensions $(D=8-10)
$, which is expected to yield further improvements in accuracy.

Looking ahead, several promising directions can be explored to further
develop and apply the algorithm. While this work focuses on the split-CTMRG
scheme for single-site PEPOs, its generalization to multi-site PEPOs is
straight forward. Additionally, to further enhance computational efficiency,
the iterative application of projectors to environment tensors may be
replaced by directly solving fixed-point equations, as discussed in Refs.~%
\cite{fishman_2018,liu_2022}.

\begin{figure}[t]
\centering
\includegraphics[width=1\linewidth]{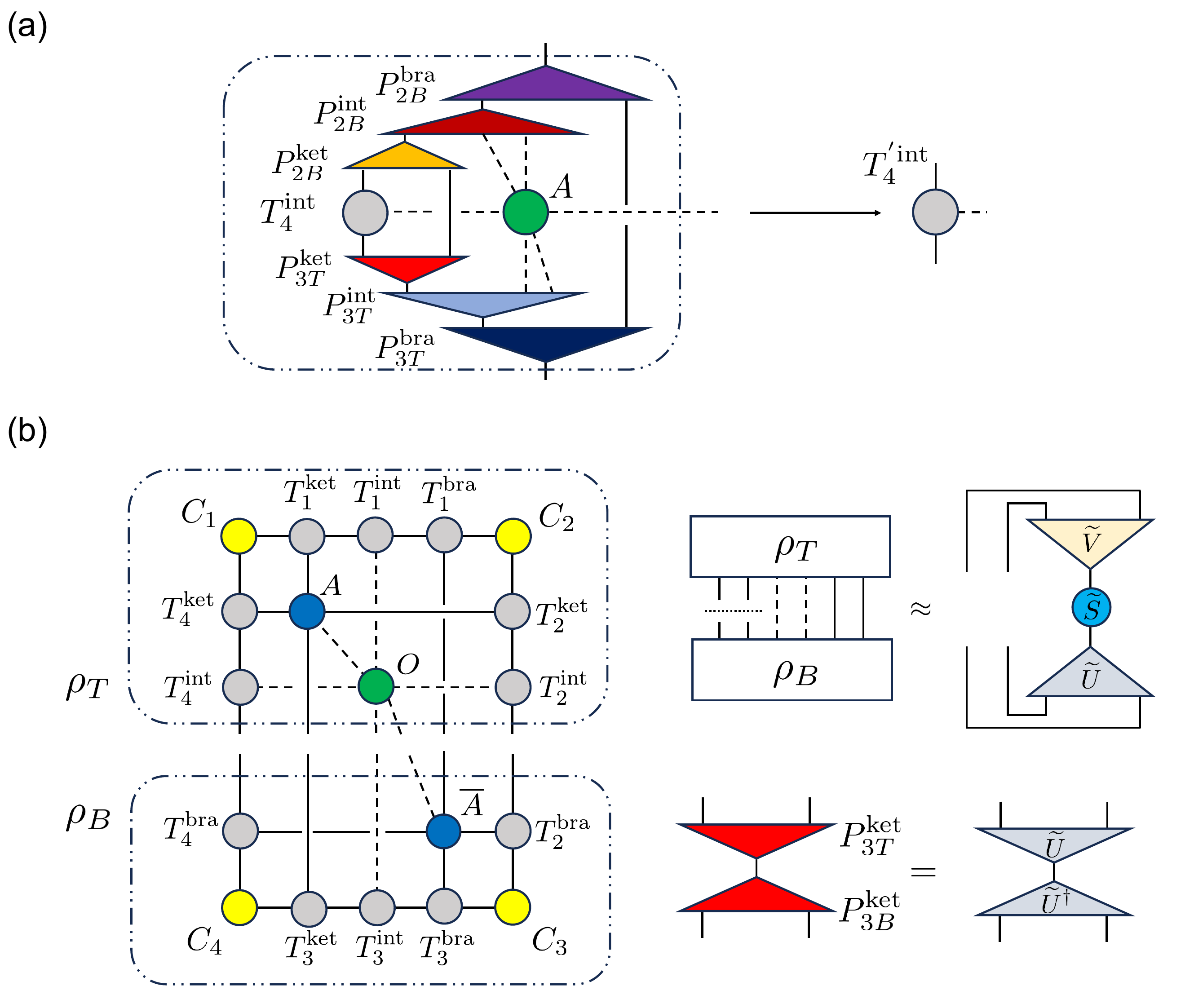}
\caption{Nine-projector scheme. (a) The absorption and truncation of $T_{4}^{%
\text{int}}$ in the nine-projector scheme. (b) The construction of projector
$P^{\text{ket}}_{3T}$ in the nine-projector scheme.}
\label{Fig::Nine-projector scheme}
\end{figure}

Although our primary focus has been on variational optimization for 3D
classical systems, the method is also applicable to 2D quantum systems,
where the Hamiltonian can be represented as a PEPO, including models with
long-range interactions \cite{pirvu2010,orourke2018}. Furthermore, for spin
systems beyond the Ising spin, it may be necessary to treat PEPOs with bond
dimensions comparable to those of PEPS, In such cases, the bonds of the
PEPOs can be truncated using separate projectors, which leads to an
alternative nine-projector scheme. As illustrated in Fig.~\ref%
{Fig::Nine-projector scheme}(a), for the renormalization of edge tensors $%
T_{4}^{\text{int}}$, three projectors $\{P_{i}^{\text{ket}},P_{i}^{\text{int}%
},P_{i}^{\text{bra}}\}$ are inserted between the vertical bonds, where the
vertical bonds of the PEPO tensor $O$ are truncated by $P_{i}^{\text{int}}$.
Since in this scheme the number of bonds truncated by each projector is
fewer than the six-projector scheme, the construction of projectors as
described in Sec.~\ref{subsection::Projectors construction}, where the
SVD applied to the non-truncated indices of
the effective environment is no longer computationally efficient. For
instance, as shown in Fig.~\ref{Fig::Nine-projector scheme}(b), when an
effective environment including a $1\times 1$ unit cell of the local tensor
set $\{A,O,\overline{A}\}$ is used to build the projectors $\{P_{3T}^{\text{%
ket}},P_{3B}^{\text{ket}}\}$, the computational cost of the SVD as given by
Eq.\eqref{eqn: projector construction} scales as $\mathcal{O}(\chi
^{3}D^{3}d^{6})$. Instead, one can employ the method proposed in Ref.\cite%
{corboz_2011}, where the SVD is applied to the bonds to be truncated [see
Fig.~\ref{Fig::Nine-projector scheme}(b)]. This would reduce the
computational cost to $\mathcal{O}(\chi ^{3}D^{3})$.

The consideration discussed above is also relevant for generalizing the
algorithms to multilayer PEPOs, such as those encountered in classical
frustrated spin systems in 3D or 2D models with multilayer structures
(e.g., multilayer or multiband superconducting systems). These extensions
will be tested in future work.

\textbf{Acknowledgments.} The authors are very grateful to Z.-Y. Xie and
S. Yang for stimulating discussions. The research is supported by the
National Key Research and Development Program of China (Grant No.
2023YFA1406400).

\textbf{Data availability.} The data that support the findings of this
article are openly available \cite{data}.

\appendix*
\section{Construction of the projectors}
\label{Appendix:Construction of projectors}

\subsection{Six-projector scheme}

\begin{figure*}[!b]
\centering
\includegraphics[width=\linewidth]{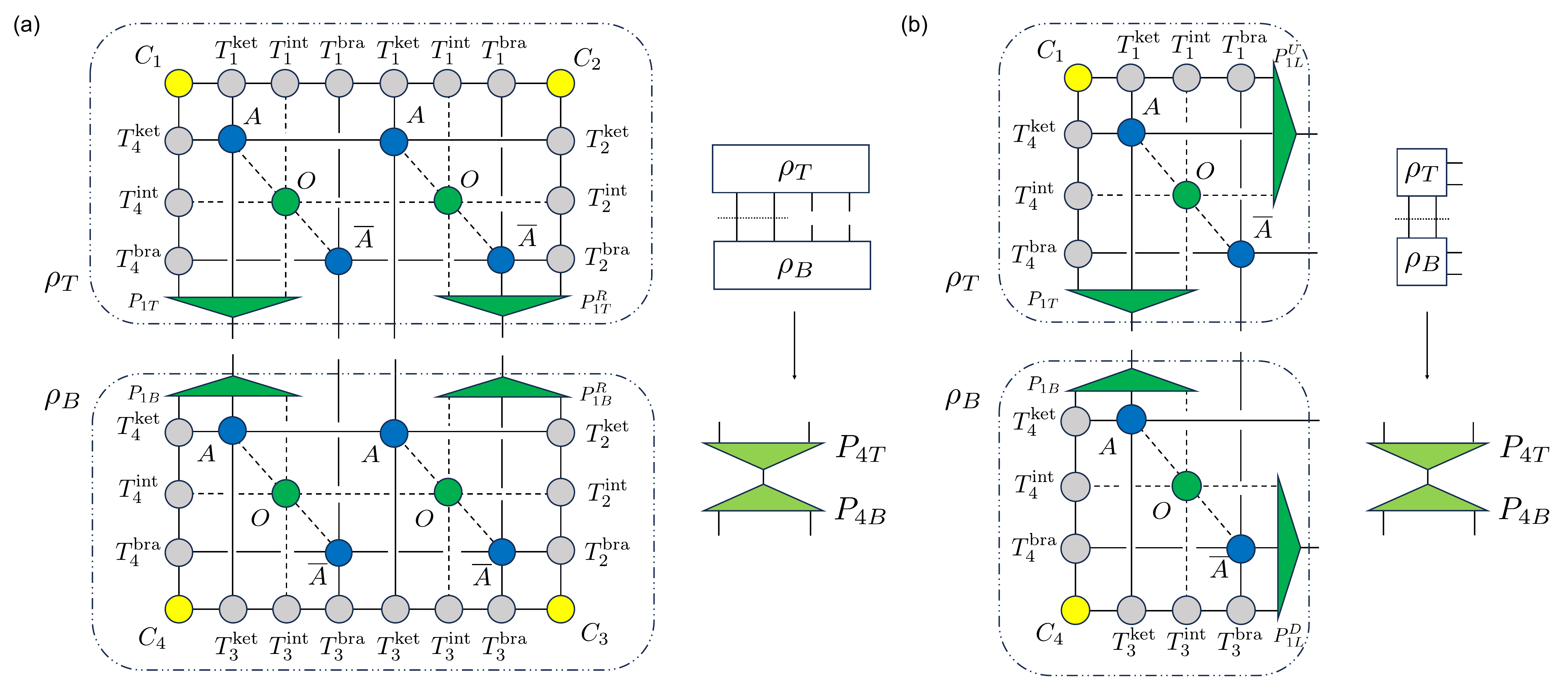}
\caption{The construction of projector 4. (a) Full projector construction. $%
P^R_{1T}$ and $P^R_{1B}$ are the projectors calculated in the right move. In
the $D_2$ symmetric split-CTMRG scheme, $P^R_{1T}=P_{1B},P^R_{1B}=P_{1T}$.
(b) Half projector construction. $P^U_{1L}$ and $P^D_{1L}$ are the
projectors calculated in the up and down move. In the $D_2$ symmetric
split-CTMRG scheme, $P^U_{1L}=(P_{1T})^T,P^D_{1L}=(P_{1B})^T$.}
\label{Fig::Projector4}
\end{figure*}

In the main text, we utilize a six-projector scheme, where the legs of the
PEPO tensor O are truncated using projectors for the ket and bra layers. The
construction of projectors 1 is illustrated in Fig.~\ref{Fig::construction
of projector 1}, and the construction of the remaining projectors is
detailed below. Since all the projectors are constructed using Eqs.~\eqref{eqn: projector construction} and \eqref{eqn: projector
construction 2}, we only provide a description of the effective environments
$\rho _{T}$ and $\rho _{B}$ for each projector. For a large effective
environment, such as those shown in Fig.~\ref{Fig::Projector4}, the optimal
contraction order varies with the choices of $d,D,\chi$ and cannot be
directly determined. Therefore, the computational cost of tensor contraction
listed below and in the main text is determined numerically, assuming $d\ll D
$, with $\chi\sim D^2$, which serves as an upper bound on the computational
cost of contraction.

\begin{figure}[t]
\centering
\includegraphics[width=1\linewidth]{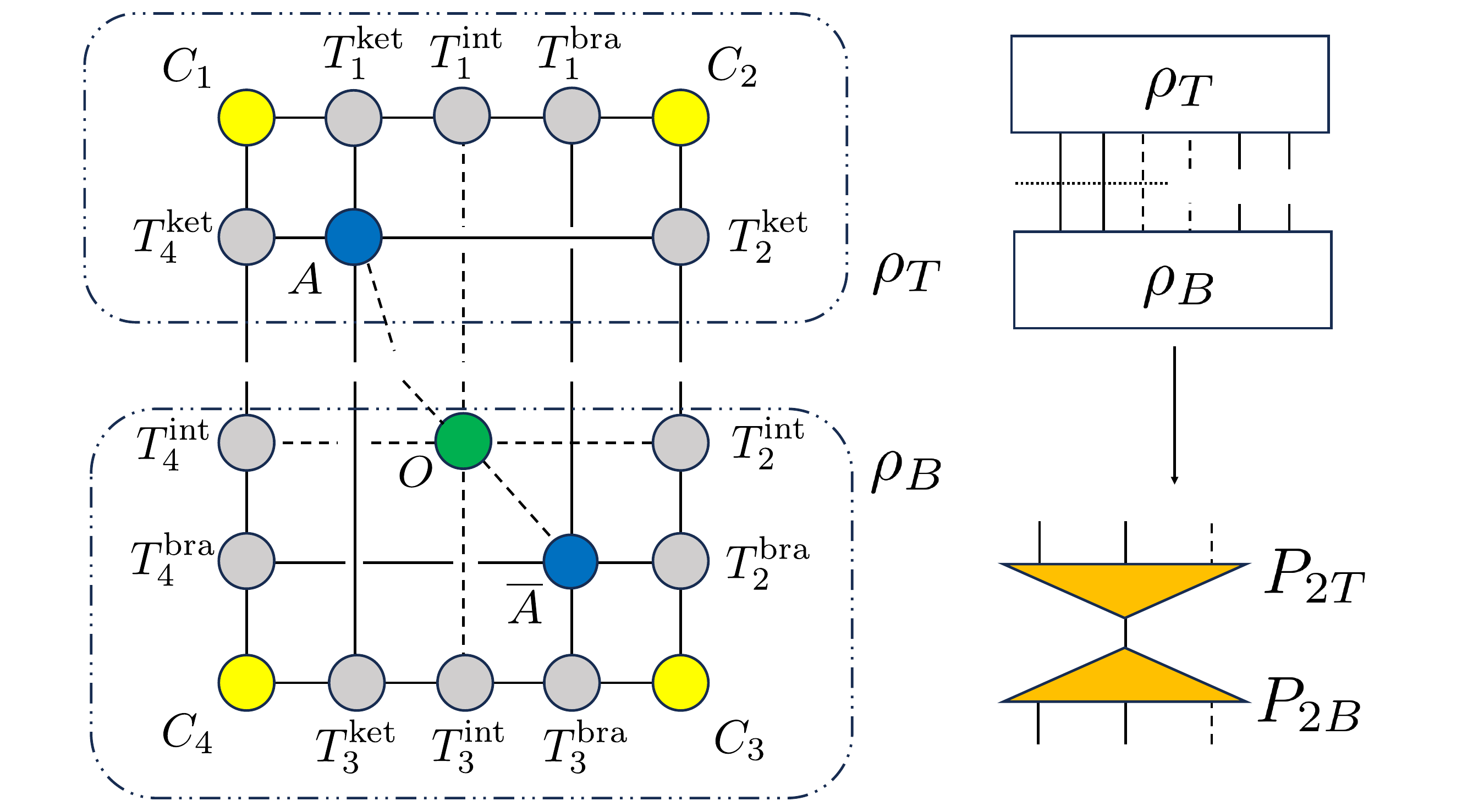}
\caption{The construction of projector 2.}
\label{Fig::Projector2}
\end{figure}

\begin{figure}[t]
\centering
\includegraphics[width=1\linewidth]{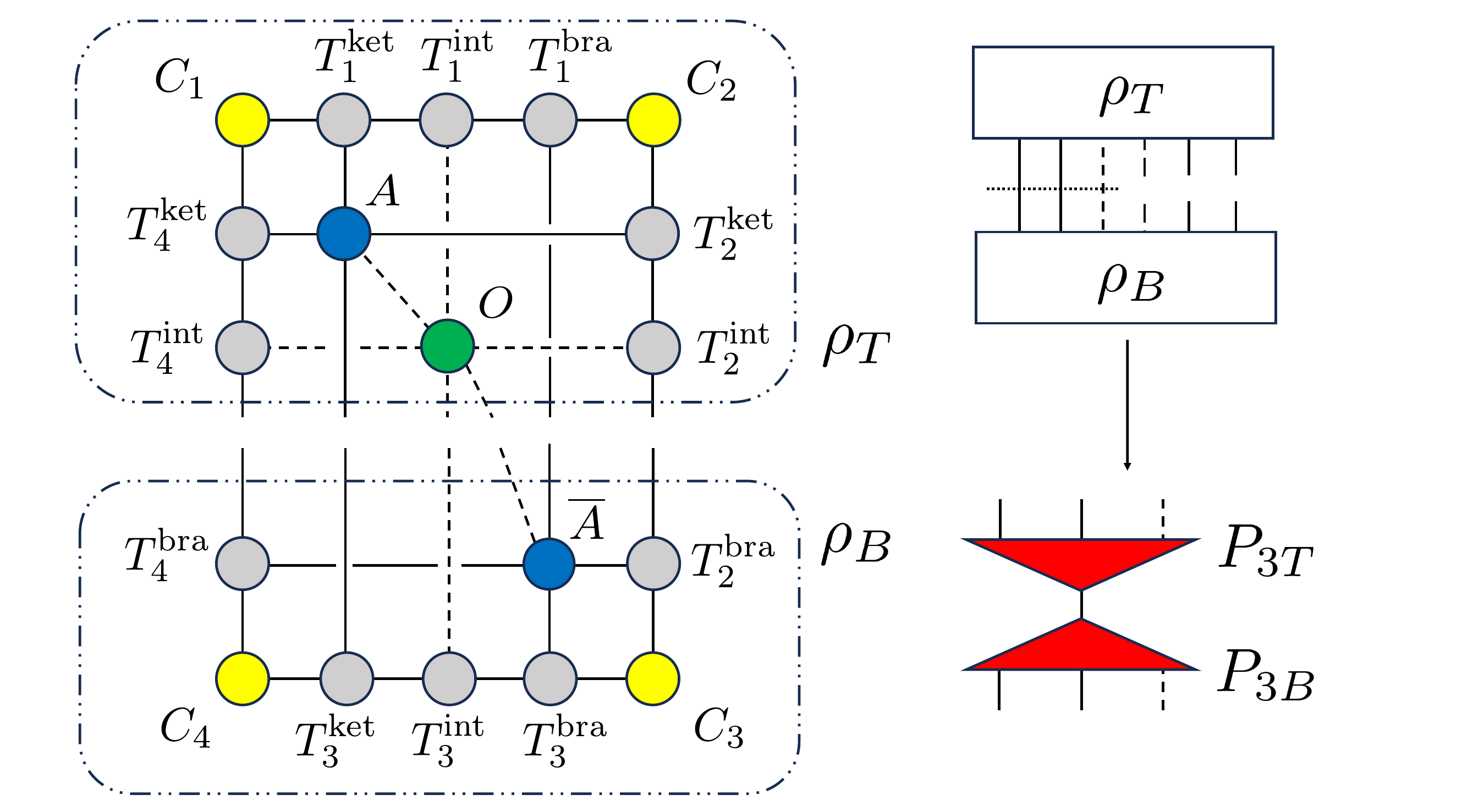}
\caption{The construction of projector 3.}
\label{Fig::Projector3}
\end{figure}

The construction of projector 2 and 3 is shown in Figs.~\ref{Fig::Projector2}
and \ref{Fig::Projector3}. The effective environment is chosen to
consist of a $1\times 1$ unitcell of the local tensor set $\{A,O,\overline{A%
}\}$, along with the surrounding environment tensors. These projectors
truncate the bond dimension from $\chi Dd$ to $\chi$. The leading
computation cost arises from tensor contraction and SVD, with a scale of
\begin{equation}
\mathcal{O}(\chi ^{3}D^{3}d^{3}).
\end{equation}
The scaling could be further reduced to $\mathcal{O}(\chi^{2}D^{4}d^{3})\sim
\mathcal{O}(D^{8}d^{3})$ using the partial SVD technique without explicitly
constructing $\rho_T$ and $\rho_B$.
The construction of projectors 4, 5, and 6 leverages projectors 1, 2 and 3 to
reduce the cost associated with SVD decompositions and tensor contraction.
For projector 4, two constructions are considered. The full projector 4 as
shown in Fig.~\ref{Fig::Projector4} (a) is built based on an effective
environment consisting of $2\times 2$ unit cells of the local tensor set $%
\{A,O,\overline{A}\}$, with the surrounding environment and local tensors
from the bra layer already renormalized by $P_{1T}$ and $P_{1B}$, while the half
projector 4, as shown in Fig.~\ref{Fig::Projector4} (b), uses $2\times 1$
unit cells of the local tensor set $\{A,O,\overline{A}\}$, with the
surrounding environment and local tensors from the bra layer renormalized by
$P_{1T},P_{1B}$ and $P_{1L}^{U},P_{1L}^{D}$. Here $P_{1L}^{U}$ and $%
P_{1L}^{D}$ represent projector 1 calculated during the up move and
down move, respectively.

For the full projector 4, the leading computation cost arises from tensor
contractions, with a scale of
\begin{equation}
\mathcal{O}(\chi ^{3}D^{4}d).
\end{equation}%
For the half projector 4, the computation cost is reduced to
\begin{equation}
\mathcal{O}(\chi ^{3}D^{3}d^{3})\sim \mathcal{O}(D^{9}d^{3}).
\end{equation}%
Although half projectors are regarded as less accurate than full projectors
in some situations \cite{corboz_2014,naumann_2025}, we don't observe
accuracy issues compared to the full projectors in our numerical benchmarks.
Therefore, in the main text, we use the computational cheaper half
projectors.

Projectors 5 and 6 are built by considering an effective environment
consisting of $1\times 2$ unit cells of the local tensor set $\{A,O,%
\overline{A}\}$, with the surrounding environment and local tensors from the
bra layer already renormalized by $\{P_{2T}$, $P_{2B},P_{2T}^{R},P_{2B}^{R}\}$ and $\{P_{3T}$, $P_{3B},P_{3T}^{R}$, $P_{3B}^{R}\}$,
respectively. Here $P_{2}^{R},P_{3}^{R}$ represent the projectors $2,3$
calculated during the right move.

\begin{figure}[h]
\centering
\includegraphics[width=1\linewidth]{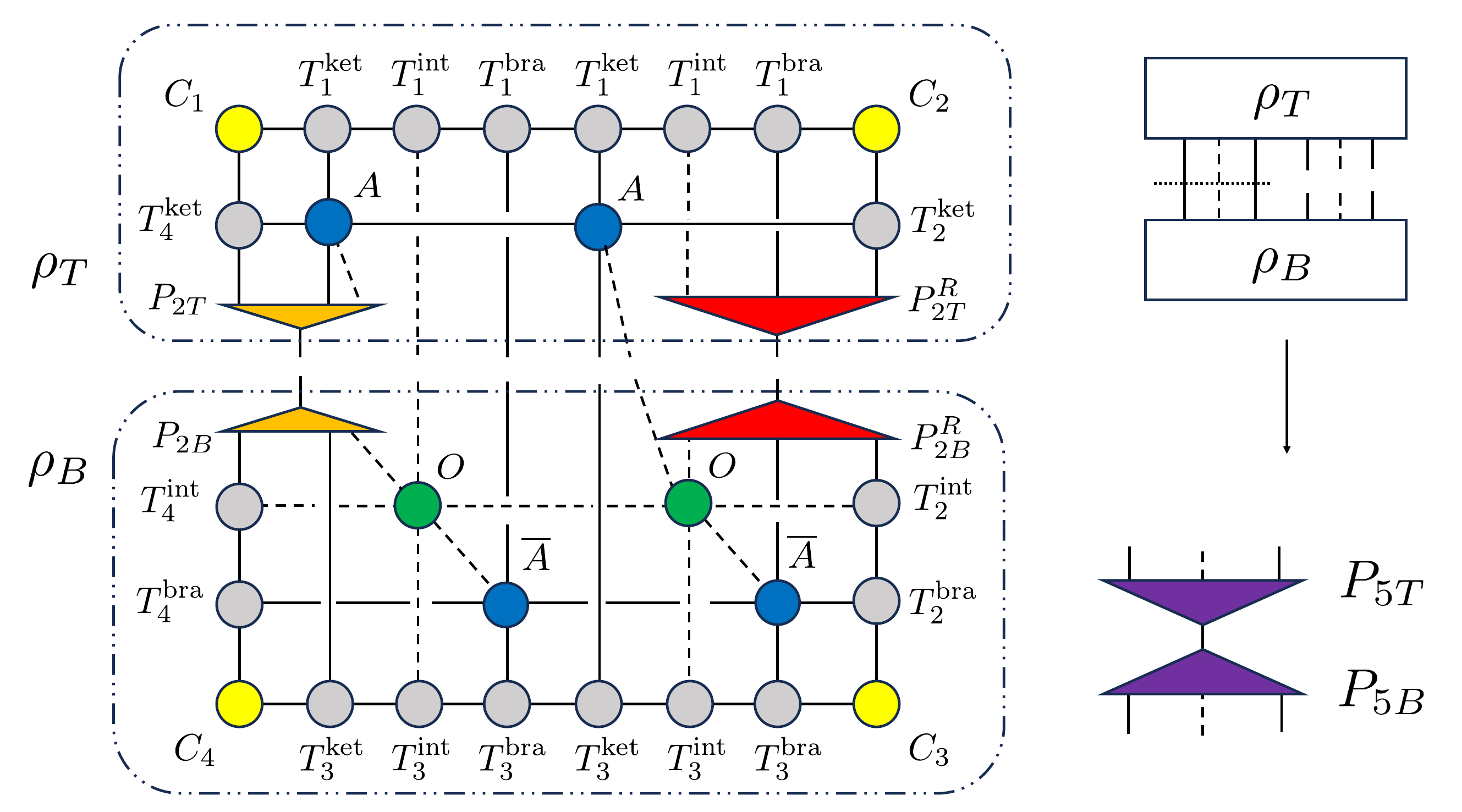}
\caption{The construction of projector 5. $P^R_{2T}$ and $P^R_{2B}$ are the
projectors calculated in the right move. In the $D_2$ symmetric split-CTMRG
scheme, $P^R_{2T}=P_{3B},P^R_{2B}=P_{3T}$. }
\label{Fig::Projector5}
\end{figure}

\begin{figure}[h]
\centering
\includegraphics[width=1\linewidth]{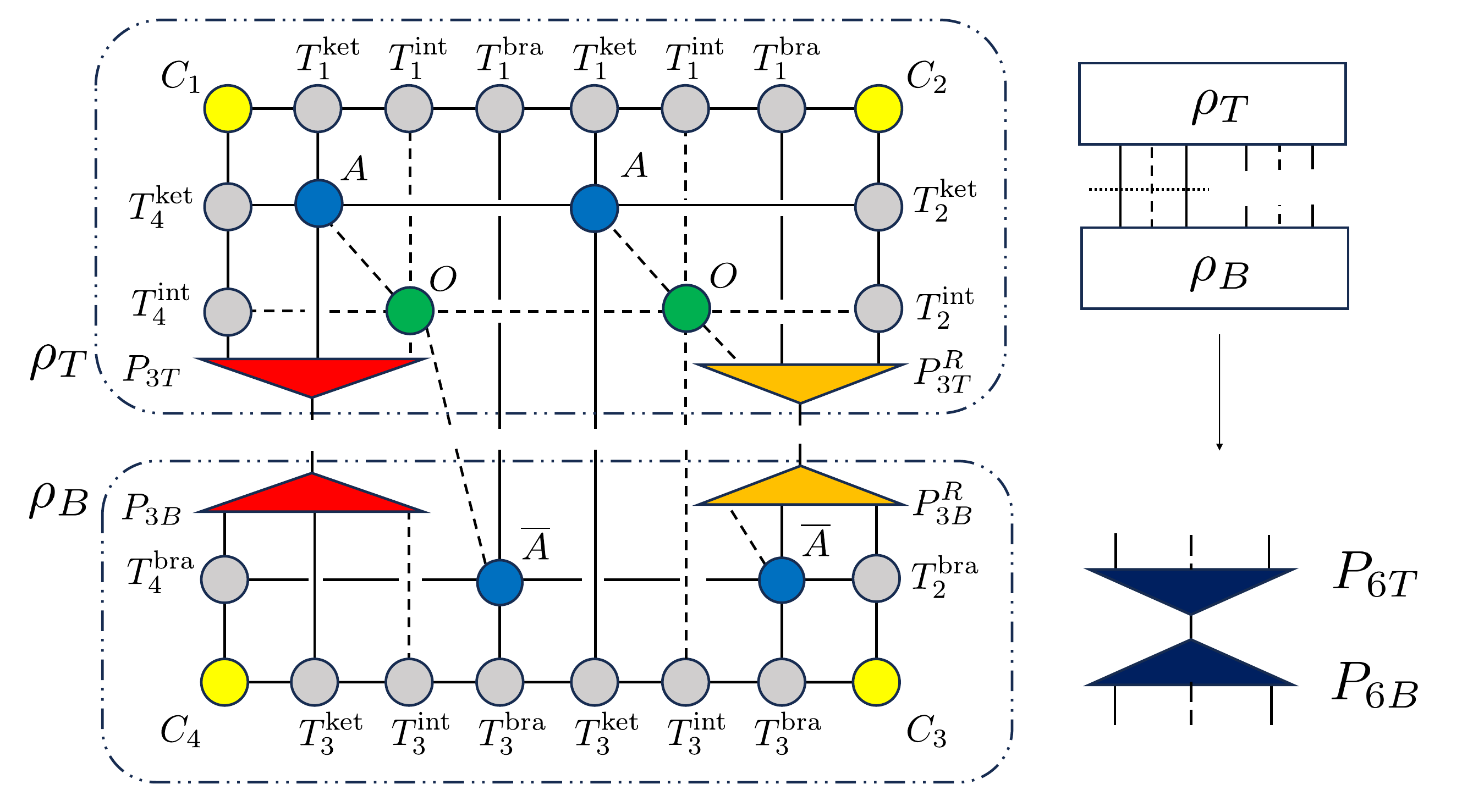}
\caption{The construction of projector 6. $P^R_{3T}$ and $P^R_{3B}$ are the
projectors calculated in the right move. In the $D_2$ symmetric split-CTMRG
scheme, $P^R_{3T}=P_{2B},P^R_{3B}=P_{2T}$.}
\label{Fig::Projector6}
\end{figure}
The leading computational cost for these projectors also arises from tensor
contractions as well as SVD, with a scale of
\begin{equation}
\mathcal{O}(\chi ^{3}D^{3}d^{3}).
\end{equation}%
The overall leading computational cost of the entire scheme is therefore
\begin{equation}
\mathcal{O}(\chi ^{3}D^{3}d^{3})\sim \mathcal{O}(D^{9}d^{3}).
\end{equation}

%

\end{document}